\documentclass[11pt, letterpaper]{article}
\usepackage{mydef2col}
\usepackage{kantlipsum,widetext}
\usepackage[T1]{fontenc}
\usepackage[autostyle=true]{csquotes}
\usepackage{listings}

\allowdisplaybreaks[3]

\usepackage[colon, authoryear, round]{natbib}
\def\calM{{\mathcal M}}
\def\calN{{\mathcal N}}

\DeclareMathOperator*{\argmin}{arg\,min}

\newcommand{\ann}{\mathrm{ANN}\mkern1mu}
\newcommand{\af}{\mathrm{af}\mkern1mu}

\newcommand*\from{\colon}

\title{Deep learning calibration of option pricing models: some pitfalls and solutions}
\author{
\authorstyle{Andrey Itkin}
\newline\newline
\institution{Tandon School of Engineering, New York University, 12 Metro Tech Center, 26th floor, Brooklyn NY 11201, USA}}

\date{\today}
\begin{document}

\maketitle

\lettrineabstract{Recent progress in the field of artificial intelligence, machine learning and also in computer industry resulted in the ongoing boom of using these techniques as applied to solving complex tasks in both science and industry. Same is, of course, true for the financial industry and mathematical finance. In this paper we consider a classical problem of mathematical finance - calibration of option pricing models to market data, as it was recently drawn some attention of the financial society in the context of deep learning and artificial neural networks. We highlight some pitfalls in the existing approaches and propose resolutions that improve both performance and accuracy of calibration. We also address a problem of no-arbitrage pricing when using a trained neural net, that is currently ignored in the literature.}

\vspace{0.5in}

\section*{Introduction}

Recent progress in the field of artificial intelligence (AI), machine learning (ML) and also in computer industry resulted in the ongoing boom of using these techniques as applied to solving complex tasks in both science and industry. Same is, of course, true for the financial industry and mathematical finance. However, as mentioned in \cite{Verge2018}, "...how fast the industry is moving, and to what end, is typically measured not just by actual product advancements and research milestones, but also by the prognostications and voiced concerns of AI leaders, futurists, academics, economists, and policymakers. AI is going to change the world — but how and when are still open questions".

In this paper we consider a classical problem of mathematical finance - calibration of option pricing models to market data, as it was recently drawn some attention of the financial society in the context of deep learning (DL) and artificial neural networks (ANN)%
\footnote{As mentioned by I. Halperin, using non-parametric methods (ANN) to calibrate parametric models is an overkill. Indeed, if one resorts to ANN, why to deal with a parametric model in the first place, since given the market data the ANN could be trained directly to this data with no model in between. In more detail, see, \cite{QLBS}}.
 Among various papers, we mention \cite{Horvath2019, Kees2019}, and also references therein. In short, the main idea of these papers is as follows. As known, given some asset pricing model the classical calibration problem consists in finding models parameters optimal in a sense that they minimize the difference (in some norm) between the prices predicted by the model and provided by the market.  In this paper for the sake of certainty we consider just the option pricing models. The idea of applying DL techniques to this problem is inspired by the fact that computing model option prices could be slow, so the whole calibration is slow. Instead, the DL calibration assumes this problem to be solved in two steps. The first step is to replace a slow pricer with its approximator by using a ANN. This ANN is trained with some in-sample set of data, so the weights of the ANN become known after this step is complete. Then a standard calibration is used where the model pricer is replaced with the trained ANN constructed at the previous step.

The advantage of this approach is that for the given model the approximating ANN could be trained offline just once, and then it could be used with any market data for the online calibration. Therefore, in the above cited papers the authors claim significant acceleration of the calibration process which could achieve several orders of magnitude (when doing such a comparison for the DL calibration, the offline time of training the ANN is ignored).

This approach seems to be attractive and efficient while still requires addressing various technical problems. A typical construction and training of the ANN is described in detail in \cite{Horvath2019}. For the general theory of using DL for asset pricing see, e.g., an extensive presentation \cite{YE2019} and references therein. However, aside of these technical details (which, certainly are very important), the approach advocated in the referenced literature has some internal pitfalls which we further discuss in this paper. Some of them are as follows. First, in our opinion the second step of the calibration process (running the global optimizer) could be eliminated as the same result could be achieved when training the ANN at the first step. This requires some more delicate consideration which we provide in the paper. Second, the ANN prices by default could not guarantee no-arbitrage, neither in-sample nor our-of sample. This is a more serious problem that requires special attention. Finally, as option Greeks are as well important as the prices themselves, one need to make sure that the ANN prices are at least $C^2$, which again requires a special consideration. However, the current approaches don't take these problems into account.

Therefore, the ultimate goal of this paper is to investigate these problems in more detail and provide some (perhaps only partial) solutions. The rest of the paper is organized as follows.

\section{Option pricing using ANN} \label{model}

In this section we give a short overview of how ANNs are used for option pricing which should be helpful to better understand the rest of the paper. For more comprehensive introduction into the subject see, e.g., recent papers \cite{Horvath2019, Kees2019,YE2019} and references therein.

\begin{figure}[!htb]
\captionsetup{format=plain}
\begin{minipage}{0.3\textwidth}
\begin{center}
\hspace*{-0.4in}
\includegraphics[totalheight=2.in]{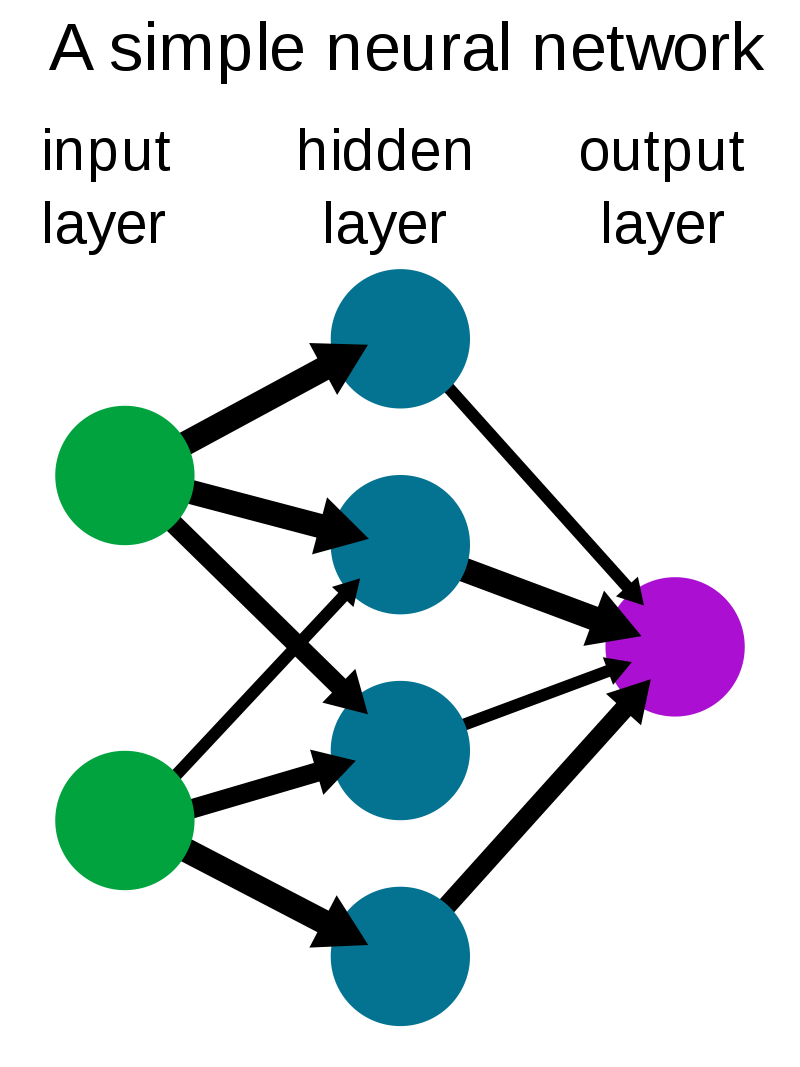}
\caption{An example of one layer feed forward ANN.}
\label{ANNfig}
\end{center}
\end{minipage}
\begin{minipage}{0.7\textwidth}
\begin{center}
\hspace*{-0.6in}
\includegraphics[totalheight=2.in]{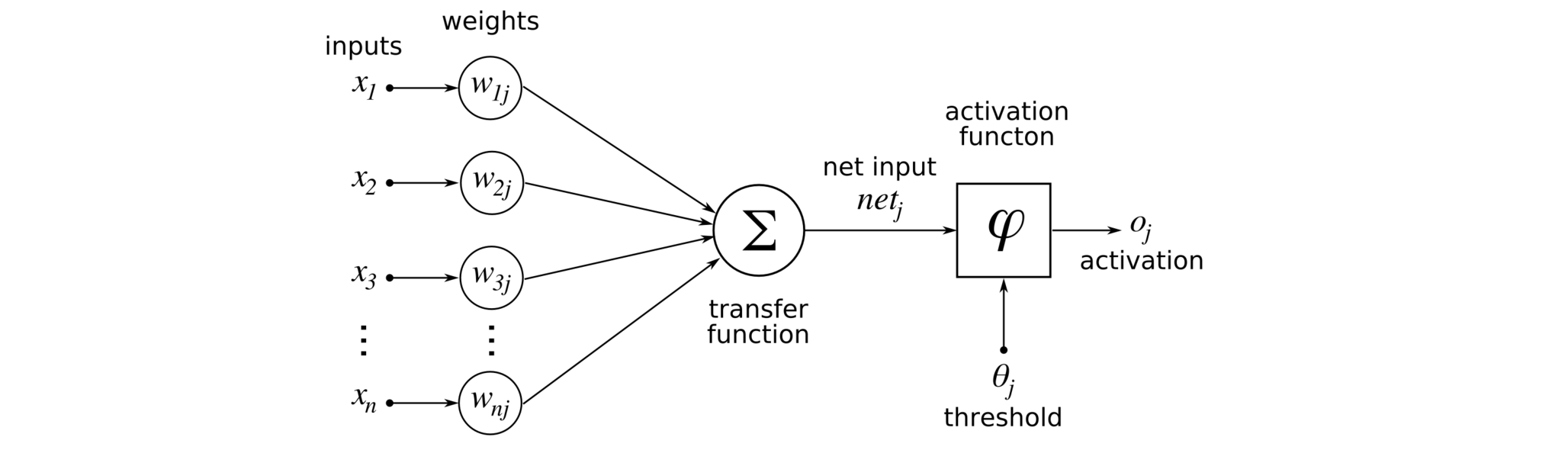}
\caption{Behavior of neuron and the activation function.}
\label{neuron}
\end{center}
\end{minipage}
\end{figure}

An artificial  neural network is a network of artificial neurons, \cite{Bishop:1995:NNP:525960}.   The connections of the neurons are modeled as weights. All inputs are modified by a weight and summed. This activity is referred as a linear combination. Finally, an activation function controls the amplitude of the output. Further we consider only a feed forward ANN which is an artificial ANN wherein connections between the nodes do not form a cycle. As such, it is different from recurrent neural networks. A simple scheme of one layer feed forward ANN is presented in Fig.~\ref{ANNfig}, and a detailed neuron behavior is highlighted in Fig.~\ref{neuron} borrowed from \cite{changhau2017}. Here $x = [x_1,...,x_m], \ x \in \mathbb{R}^m$ are the ANN inputs, $w_j = [w_{i,j},...,w_{i,m}], \ i \in [1,m], w_{i,j} \in \mathbb{R}^m$ are the network weights corresponding to the hidden layer $j$ and connecting nodes $1,...,m$ to the next layer, and $o_j = y$ is the output. The function $\Sigma$ is often chosen in the form $\Sigma = b_j + \sum_{k=1}^m w_{k,j} x_k$ where $b_j \in \mathbb{R}$ denotes the bias term. There exists a wide range of activation functions used in practice, again in more detail, see, e.g.,  \cite{changhau2017}.

Suppose, we have some function $y=F(x)$ and need to construct a feed forward ANN: $y=F_{\ann}(x,w)$ approximating this function in some optimal sense. For doing so, the ANN should be trained by using a set of input data by running some optimization. This optimization takes as an input a sequence of training examples $(x_1,y_1),\ldots ,(x_m,y_m)$ and produces a sequence of weights $w_{i,j}$ optimal in a sense that for the given training they minimize the difference
$\sum_i \|F(x_i) - F_{\ann}(x_i)\|$ taken under some appropriate norm, e.g., $L^2$.

We now want to apply this ANN approximation to the option pricing problem. Consider a financial model $\mathcal{M}$ with parameters of the model $p_1,\ldots,p_n, \ p_i \in \mathbb{R}, n \in \mathbb{Z}, n \ge 1$ which provides prices of some financial instruments given the input data $\theta_1,\ldots,\theta_l, \ l \in \mathbb{Z}, l \ge 1$. For example, the celebrated Black-Scholes model has one model parameter $\sigma$ - the volatility of the stock, and 5 input parameters $S,K,T,r,q$ which are the stock price, the strike, the time to maturity, the free interest rate and the continuous dividend. The DL approach for option pricing basically assumes that the ANN can be used as a universal approximator $\mathbb{R}^{n+l}  \to \mathbb{R}$, i.e. given a vector of the input data $\bm{\theta}$ and a vector of the values of the model  parameters $\bm{p}$ it provides a unique option price, e.g., the Call option price $C(\bm{\theta},\bm{p})$.

So why this it useful? Recent significant progress in using ANNs for scientific and industrial applications could be explained, perhaps,  by several factors. First, the amount of available input/output data rapidly grew within last decades, hence training of the ANN could be done more accurately. Second, modern computers continue to be more and more powerful, and even personal computers are now equipped with multicore CPUs, GPU and hundreds GBs of RAM. Therefore, tedious calculations could be performed fast. Third, various free software packages and APIs such as Tensorflow of Google, Keras of MIT, etc. had become available for the society, which significantly simplified programming of ANNs. Overall, this progress makes it possible solving complex multi-dimensional problems in a realistic time.

Having this in mind, let us mention two possible approaches for training the ANN pricer. The first one is simple, and assumes that the ANN pricer is trained given a single set of inputs, $\bm{\theta}_0$. So it provides a pricing map $\calM_C\from C(\bm{\theta}_0,\bm{p})  \to  C_{\ann}(\bm{\theta}_0,\bm{p})$ where $\bm{p} = [\bm{p}_1,...,\bm{p}_N]$ is the training set of vectors of the model parameters. Again, for the Black-Scholes model this is simply $\bm{p} = [\sigma_1,...,\sigma_N]$. From the practical point of view this approach is not feasible as the ANN should be re-trained for every new set of  inputs $\bm{\theta}$. For instance, if the stock price changes in the future, this ANN pricer should be re-trained.

The second approach would be to combine inputs and parameters of the model into a single vector $\bm{\xi} = [\bm{\theta},\bm{p}]$ and train the ANN as $\bm{\xi}  \to  C_{\ann}(\bm{\xi})$. This requires more intensive computational work but eliminates any re-training. Hence, the ANN pricer could be trained once and forever. Based on the notes in the previous paragraph about the efficiency and power of the modern computer systems, and having in mind that this job is done offline, this task is perfectly doable. Thus, in "old school language" the ANN pricer could be treated as a tensor lookup table created by using nonlinear multi-dimensional regressions.

A typical example of implementation of this algorithm in python using Keras could be found, e.g., in \cite{VarmaDas2018}. After the ANN is trained, computation of option prices on the ANN is very fast both in-sample and out-of sample. This is especially pronounced for models that require using Monte Carlo pricers which are very slow.

\section{ANN and option Greeks} \label{ANNGreeks}

In addition to option prices traders, front office and risk management quants also need option Greeks. Therefore, the ANN prices must be, at least $C^1$, or better $C^2$. From this prospective in \cite{Horvath2019} an important theorem from \cite{Hornik1990} is cited, which is also reproduced below for the reference.

\begin{theorem}[\cite{Hornik1990}]
Let $\mathcal{N}^\sigma_{d_0,d_1}$ be the set of ANN with activation function $\sigma: \mathbb{R} \mapsto \mathbb{R}$ , input dimension $d_0 \in \mathbb{N}$ and output dimension $d_1 \in \mathbb{N}$. Let $F \in C^n$, and $F_{\ann}: \mathbb{R}^{d_0} \to \mathbb{R}$.  Then, if the (non-constant) activation function is $\sigma \in C^n(\mathbb{R})$, then $\mathcal{N}^\sigma_{d_0,d_1}$ arbitrarily approximates $F$ and all its derivatives up to order $n$.
\end{theorem}
Thus, differentiability of the activation function is important for option pricing.

If the activation function is not applied at all, the output signal becomes a simple linear function. Linear functions are only single-grade polynomials, therefore a non-activated neural network will act as a linear regression with limited learning power, \cite{Walia2017}.  As ANN are designed as universal function approximators, they are intended to learn any function. Thanks to the non-linear activation functions, stronger learning of networks can be achieved. However, not any activation function has good properties together with the existence of first two derivatives. For instance, ReLu could be a good candidate, but it suffers from the problem of vanishing gradients. One can decide to use Leaky ReLU as a solution, but it is not $C^1$ at the origin. So the choice of the activation function could be critical. From this prospective, ELU activation function
\[ R(z) =
\begin{cases}
z & z > 0, \\
\alpha(e^z-1) & z \le 0, \\
\end{cases}
\]
\noindent with the hyperparameter $\alpha = 1$  could be a choice , so the first derivative of the ELU is smooth. However, the second derivative jumps at $z=0$. This function also has two important properties: i) it produces a zero-centered distribution, which can make the training faster, and ii) it provides one-sided saturation which leads to a better convergence, see, e.g., \cite{Kathuria2018}. However, if the output of the last ANN layer is the option price, ELU cannot be used as it doesn't guarantee the positivity of the price. Therefore, one has to use either another activation function for the last layer, or scale the price, so the scaled price could become negative.

In our test example we generated 300000 random vectors $\bm{\xi} = [S,K,T,r,q,\sigma]$, and then computed the Call option price using the Black-Scholes model. We then left only those prices that obey $0.001 \le C \le 10$, which provided 259012 samples. 80\% of them where used as a training set, and the remaining 20\% - as a prediction set. We built a feed forward ANN with 128 nodes and 4 layers, so the total number of trainable parameters was 33921. For training the ANN we used either Rmsprop or Adam optimizer with MSE loss function, 15 epochs and batch size 64.  The activation functions for the layers were as follows:
\begin{enumerate}
\item LeakyReLU with $\alpha = 1$. Then this function is $C^2$.
\item Custom activation function which is a modified ELU (MELU):
\[ R(z) =
\begin{cases}
\frac{\frac{1}{2}z^2 + a z}{z + b} & z > 0, \\
\alpha(e^z-1) & z \le 0, \\
\end{cases}
\qquad a = 1 - 2 \alpha, \quad b = -2 + \frac{1}{\alpha}.
\]
It could be verified that  $R(z) \in C^2$, and $R'(0) = R''(0) = \alpha$.

\item Same as in 2.
\item Softplus function minus 0.5. This function is also $C^2$.
\end{enumerate}
Thus, all activation functions in use are $C^2$, therefore same is true for the whole ANN approximation. In our test we choose $\alpha = 0.49$, therefore $\mathrm{MELU}(x) > -0.5$ . Accordingly, we scale the inputs: $S \mapsto S/K, \ C \mapsto C/K - \min_{\bm{\xi}}(C/K) - 0.5$.

\begin{figure}[!htb]
\captionsetup{format=plain}
\begin{minipage}{0.4\textwidth}
\centering
\hspace*{-0.2in}
\includegraphics[totalheight=2.2in]{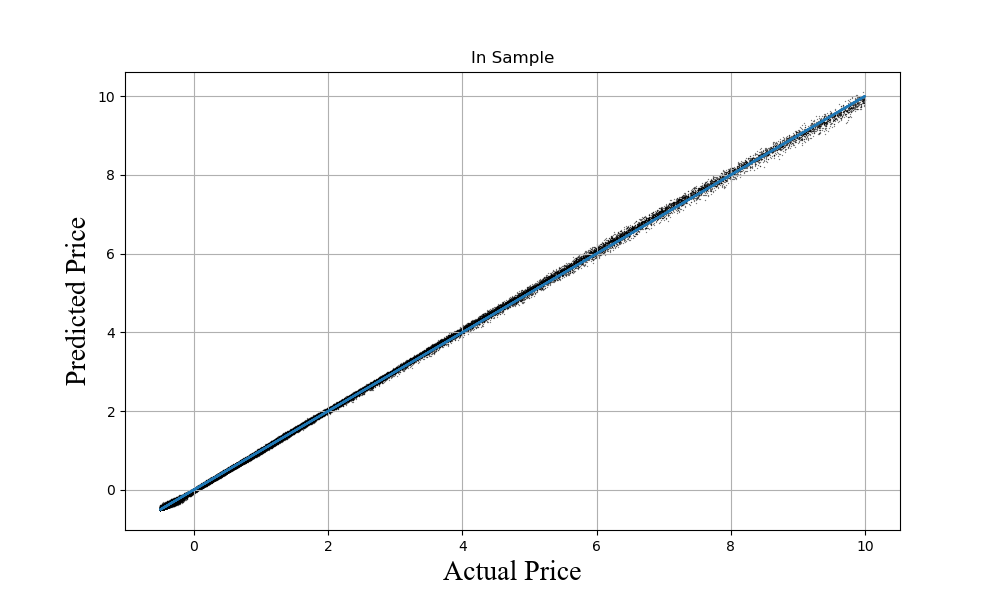}
\caption{In-sample ANN Black-Scholes scaled Call option price vs the Black-Scholes scaled Call option price.}
\label{BSpriceIn}
\end{minipage}
\hspace{0.1\textwidth}
\begin{minipage}{0.4\textwidth}
\centering
\hspace*{-0.2in}
\includegraphics[totalheight=2.2in]{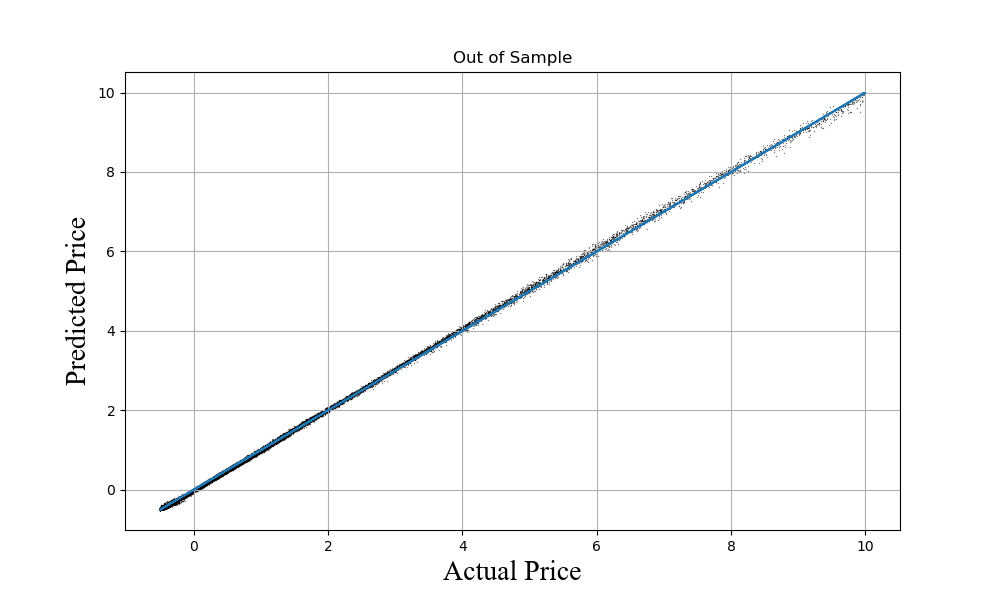}
\caption{Out-of sample ANN Black-Scholes scaled Call option price vs the Black-Scholes scaled Call option price.}
\label{BSpriceOut}
\end{minipage}
\end{figure}

\begin{figure}[!htb]
\captionsetup{format=plain}
\begin{minipage}{0.4\textwidth}
\centering
\hspace*{-0.2in}
\includegraphics[totalheight=2.2in]{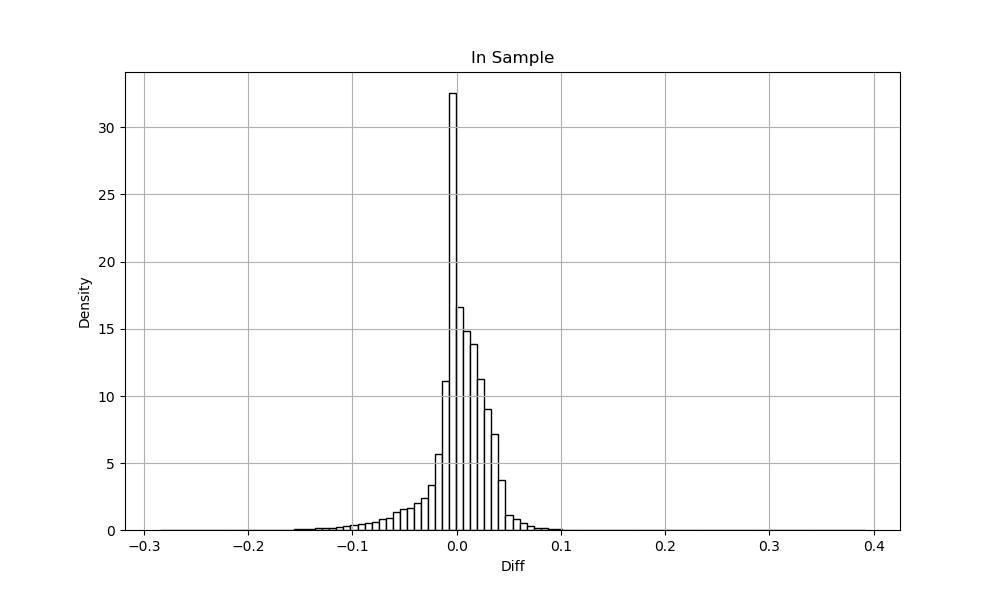}
\caption{In-sample $\delta$ error density of the Black-Scholes scaled Call option price.}
\label{BSdenIn}
\end{minipage}
\hspace{0.1\textwidth}
\begin{minipage}{0.4\textwidth}
\centering
\hspace*{-0.2in}
\includegraphics[totalheight=2.2in]{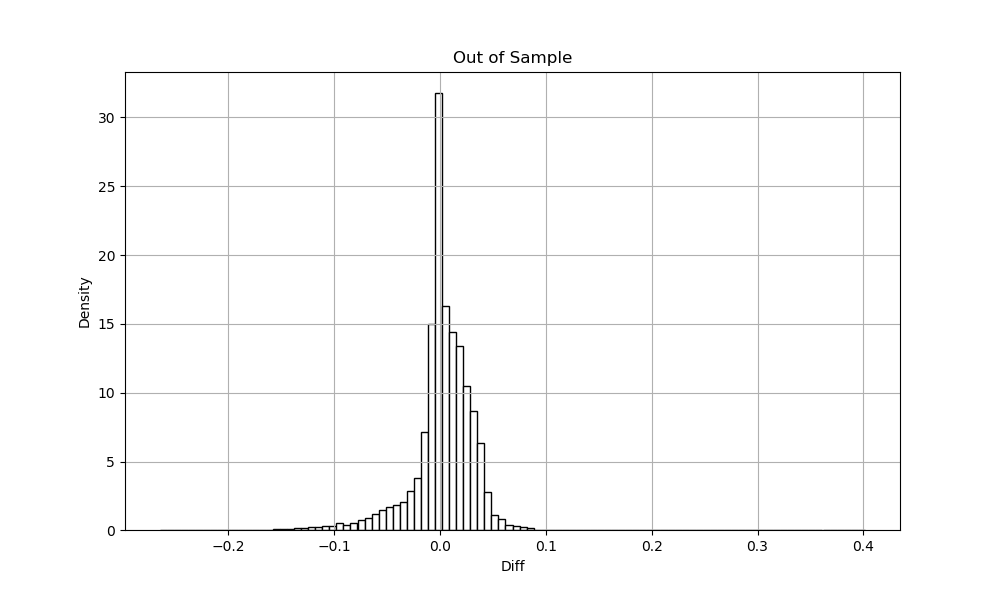}
\caption{Out-of sample $\delta$ error density of the Black-Scholes scaled Call option price.}
\label{BSdenOut}
\end{minipage}
\end{figure}

The results of this test are presented in Fig.~\ref{BSpriceIn},\ref{BSpriceOut} for in-sample and out-of sample inputs. And Fig.~\ref{BSdenIn},\ref{BSdenOut} represent the density of the difference $\delta = y - y_{\ann}$. The code ran on PC with Intel i7-4790 CPU (8 Cores, 3.6GHz), and one epoch took about 15 seconds.  The MSE both in-sample and out-of  sample is 9.7 bps, and Mean percent error is 0.1 after 15 epochs.

In case the predicted price exactly matches the actual price, all scattering points on the price graphs should coincide with
the line $y(x)=x$ which is also plotted in the figures as a solid line.

Tensorflow package also makes it possible to obtain derivatives of the option prices with respect to all input and model parameters in $\bm{\xi}$. Therefore, first order and second option Greeks are available on the ANN,  and are continuous according to the construction of activation functions.

Knowing derivatives on the ANN is also necessary to discuss no-arbitrage conditions, a topic which the next Section is dedicated to.

\section{ANN option prices and no-arbitrage}

It is known from the option pricing theory, e.g., \cite{CoxRubinstein1985,CarrMadan2005}, that the necessary and sufficient conditions for the Call option prices $ C = C(S,K,T,r,q)$ to be arbitrage-free read
\begin{equation} \label{NA}
\fp{C}{T} > 0, \quad \fp{C}{K} < 0, \quad \sop{C}{K} > 0.
\end{equation}
A discrete version of the conditions \eqref{NA} is as follows. Given three Call option prices $C(K_1), C(K_2) ,CP(K_3)$ for three strikes $K_1 < K_2 < K_3$, and all other parameters ($S, T,r,q$) being the same, the necessary and sufficient conditions for this system to be arbitrage-free read
\begin{align} \label{noarb}
C(K_3) &> 0, \qquad C(K_2) > C(K_3), \\
(K_3 - K_2)C(K_1) &- (K_3 - K_1)C(K_2) + (K_2 - K_1)C(K_3) > 0. \nonumber
\end{align}
The second condition represents a "vertical spread", and the third one - a "butterfly spread". Also, given two Call option prices $C(T_1), C(T_2)$ for two maturities $T_1 < T_2$, and all other parameters ($S, K,r,q$) being the same, the third necessary and sufficient condition for this system to be arbitrage-free read (a "calendar spread")
\begin{equation} \label{noarb}
 C(T_2) > C(T_1).
\end{equation}
Suppose that the Call option prices are computed by using the trained ANN, similar to how this is done in Section~\ref{ANNGreeks}. By default we cannot guarantee that the conditions in \eqref{NA} are satisfied (despite if the ANN is well trained they could be almost satisfied, or even fully satisfied) . This is because when the ANN is trained the optimizer knows nothing about these conditions. Thus, if we want the in-sample ANN prices to obey the no-arbitrage conditions, we need to replace unconditional optimization
with its conditional version. This is similar, for instance, to how a no-arbitrage parametric implied volatility surface is constructed by running a global constrained optimization, see e.g., \cite{ItkinSigmoid2015}.

As applied to the ANN this approach can be implemented as follows. Any constraints (including the no-arbitrage constraints) can be imposed as soft constraints by adding new (penalty) terms to the loss (objective) function that is minimized during training, see \cite{SmithCoit1996} for a general introduction to penalty functions.  It is known that this approach has two drawbacks. First, it makes it necessary to wisely choose the relative importance of the different terms in the loss function to let the optimizer to converge. Sometimes this could not be easy. Second, in contrast to hard constraints, soft constraints doesn't fully guarantee that the constraints are exactly satisfies, but only with some accuracy. The latter issue could be improved by running optimization iteratively, and using the result found at the previous iteration as the initial guess for the next one. In doing so, the soft constraints at each iteration are adjusted to make them being closer to the hard constraints.

However, as the authors of \cite{HardConstraintsDL2017} surprisingly observed, imposing soft constraints instead of hard ones yields even better results, while being far less computationally demanding when applied  in the framework of DL. Therefore, in this paper we discuss just the soft constraints, while further investigation on this topic would be definitely encouraged.

With this arguments in mind, suppose that at every layer $j$ ANN training of weights $w$ is provided by using the MSE loss function
\begin{equation} \label{MSE1}
L = \argmin_{w} \sum_i \sum_j \left[C(\bm{\xi}_i) - C_{\ann}(\bm{\xi}_i, w_{i,j}) \right]^2
\end{equation}
This loss function can be penalized by the following constraints
\begin{align} \label{MSEnext}
L_c &= \argmin_{w} \sum_i \sum_j \Bigg\{\left[C(\bm{\xi}_i) - C_{\ann}(\bm{\xi}_i, w_{i,j}) \right]^2
+ \Phi_{\lambda_1, m_1}  \left( -\sop{C_{\ann}(\bm{\xi}_i, w_{i,j})}{K} \right) \\
&+ \Phi_{\lambda_2, m_2}  \left( -\fp{C_{\ann}(\bm{\xi}_i, w_{i,j})}{T} \right)
+ \Phi_{\lambda_3, m_3} \left( \fp{C_{\ann}(\bm{\xi}_i, w_{i,j})}{K} \right) \Bigg\},  \nonumber \\
\Phi_{\lambda, m}(x) &=
\begin{cases}
0, & x < 0 \\
\lambda x^m, & x \ge 0,
\end{cases}
\nonumber
\end{align}
\noindent where $T, K \in \bm{\xi}_i$, and $\lambda \in \mathbb{R}, \ m \in \mathbb{Z}$ are some positive constants to be appropriately chosen. These constants control how strongly the constraint will be enforced. The penalty functions modify the original objective function, so that if any inequality constraint is violated, a penalty is invoked. And if all constraints are satisfied, there is no penalty.

The penalty function $\Phi_{\lambda, m}(x)$ is $m-1$ is $m-1$ times differentiable. Therefore, choosing $m>2$ would be helpful so it should not cause any trouble in an optimization algorithm which relies on first or second derivatives.

Obviously, if the constants $\lambda$  are too large, the optimization problem transforms to minimizing the constraints while the loss function is almost ignored. The opposite case is when these constants are small. Therefore, they have to be taken so to not make the penalty terms either negligible or too significant. From this prospective it is useful to normalize \eqref{MSEnext} in the following way
\begin{align} \label{MSEnorm}
L_c &= \argmin_{w} \sum_i \sum_j \left[C(\bm{\xi}_i) - C_{\ann}(\bm{\xi}_i, w_{i,j}) \right]^2 + \mathcal{P}, \\
\mathcal{P} &=  \sum_i \sum_j \Bigg\{\Phi_{\lambda_1, m_1} \left(-K^2 \sop{C_{\ann}(\bm{\xi}_i, w_{i,j})}{K} \right) +   \Phi_{\lambda_2, m_2}  \left( - T \fp{C_{\ann}(\bm{\xi}_i, w_{i,j})}{T} \right) \nonumber \\
&+   \Phi_{\lambda_3, m_3} \left(K \fp{C_{\ann}(\bm{\xi}_i, w_{i,j})}{K} \right) \Bigg\}.  \nonumber
\end{align}

Technically, this can be implemented as a custom loss function, for instance, in Keras which provides an appropriate interface. Also the necessary gradients could be obtained inside this custom loss function. Therefore, implementation of $L_c$ defined in \eqref{MSEnorm} is fully supported.

To test this approach we again use the example considered in Section~\ref{ANNGreeks}. As a measure of arbitrage we compute the entire penalty term ${\mathcal P}$ in \eqref{MSEnorm} assuming $\lambda_1 = \lambda_2 = \lambda_3 = 1$ and $m_1 = m_2 = m_3 = 0$ which is further denoted as ${\mathcal P}_{1,0}$. Thus, in the case of no-arbitrage ${\mathcal P} = 0$. The more is the value of the term, the more arbitrage prices the set predicted by the trained ANN contains. In our experiments we choose $m_1 = m_2 = m_3 = 4$, and vary $\lambda_i, \ i=1,2,3$. The results are presented in Table~\ref{arbTab}. The case $\lambda_1 = \lambda_2 = \lambda_3 = 0$ corresponds to no penalty.

\begin{figure}[!htb]
\captionsetup{format=plain}
\begin{minipage}{0.4\textwidth}
\centering
\hspace*{-0.2in}
\includegraphics[totalheight=2.2in]{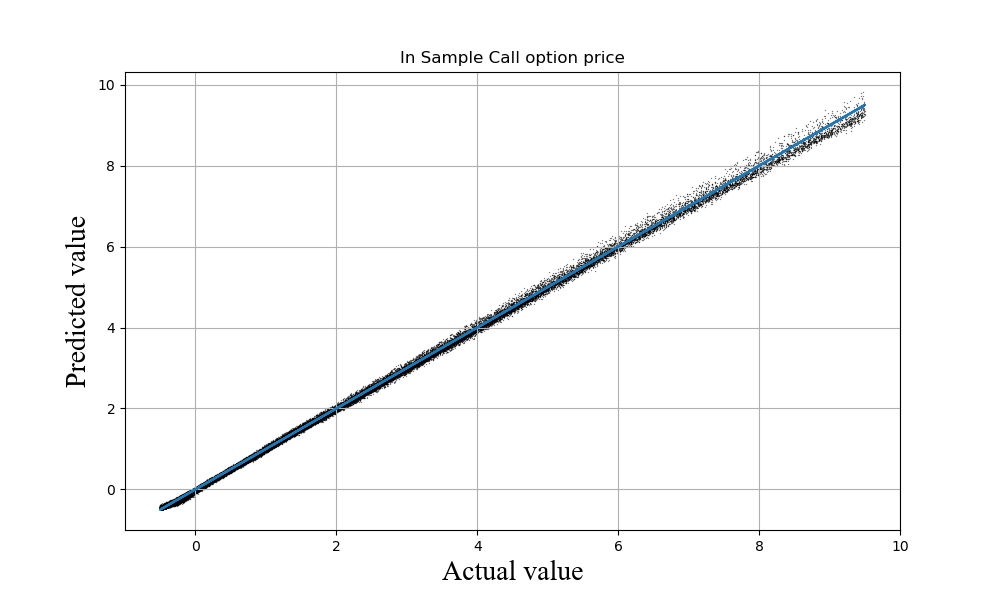}
\caption{In-sample constrained ANN Black-Scholes scaled Call option price vs the Black-Scholes scaled Call option price.}
\label{BSpriceInNA}
\end{minipage}
\hspace{0.1\textwidth}
\begin{minipage}{0.4\textwidth}
\centering
\hspace*{-0.2in}
\includegraphics[totalheight=2.2in]{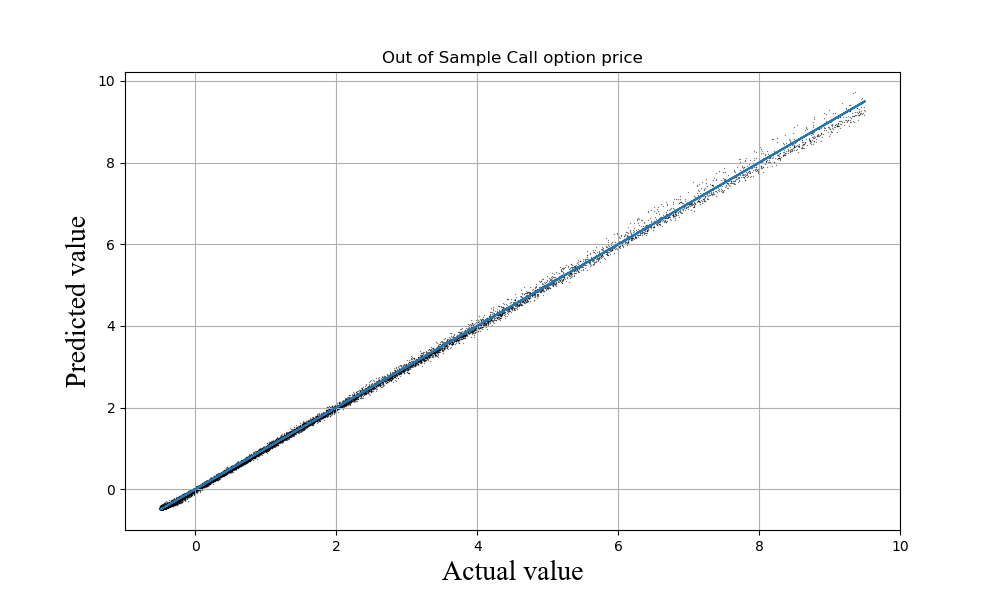}
\caption{Out-of sample  constrained ANN Black-Scholes scaled Call option price vs the Black-Scholes scaled Call option price.}
\label{BSpriceOutNA}
\end{minipage}
\end{figure}

\begin{figure}[!htb]
\captionsetup{format=plain}
\begin{minipage}{0.4\textwidth}
\centering
\hspace*{-0.2in}
\includegraphics[totalheight=2.2in]{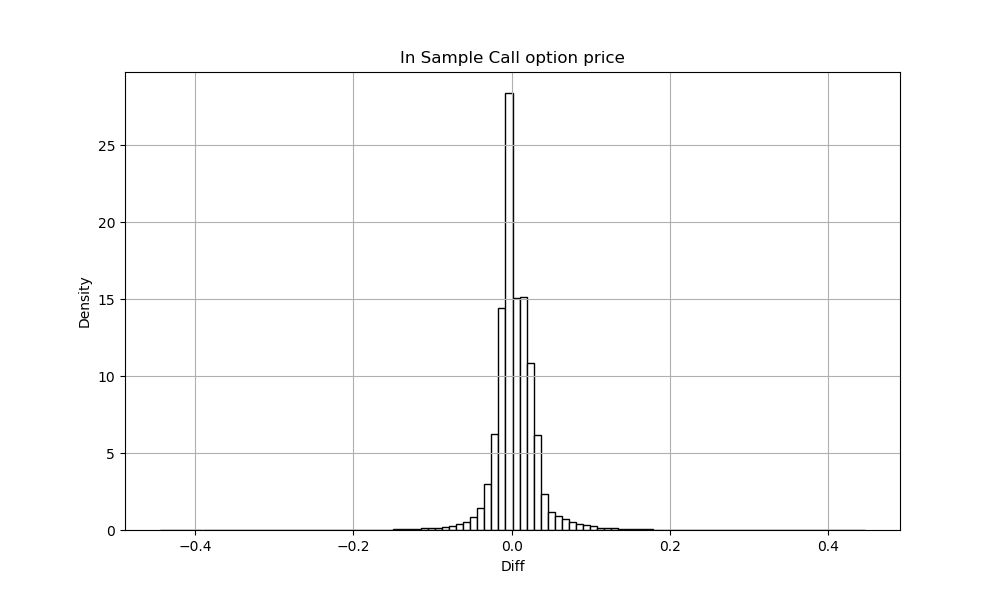}
\caption{In-sample $\delta$ error density of the Black-Scholes scaled Call option price predicted by the constrained ANN.}
\label{BSdenInNA}
\end{minipage}
\hspace{0.1\textwidth}
\begin{minipage}{0.4\textwidth}
\centering
\hspace*{-0.2in}
\includegraphics[totalheight=2.2in]{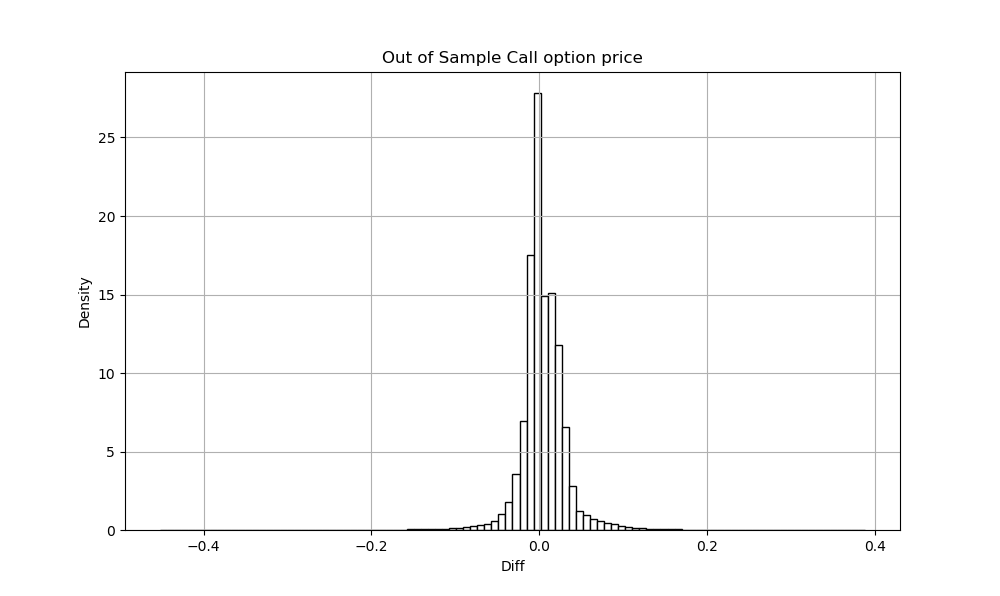}
\caption{Out-of sample $\delta$ error density of the Black-Scholes scaled Call option price  predicted by the constrained ANN.}
\label{BSdenOutNA}
\end{minipage}
\end{figure}

\begin{table}[!htb]
\begin{center}
\begin{tabular}{|c|c|c|r|r|r|r|r|}
\hline
\rowcolor{cyan}
$\lambda_1$ & $\lambda_2$ & $\lambda_3$ & \mbox{In-sample} & \mbox{MSE, bps} & \mbox{Mean \% error}
& \mbox{Out-of sample} & \mbox{MSE, bps} \\
\hline
0 & 0 & 0 & 1008550 & 6.50 & 0.091 & 248432 &  6.48\\
\hline
1 & 1 & 1 & 421 &  10.32 & 0.115 &  101 & 10.38 \\
\hline
10 & 10 & 10 & 105 & 11.95 & 0.124 &  25 &  12.15\\
\hline
50 & 50 & 50 & 36 & 12.17 & 0.125 & 9 & 12.34 \\
\hline
100 & 100 & 100 & 25 & 11.26 & 0.120 & 6 & 11.43 \\
\hline
\end{tabular}
\caption{Penalty term ${\mathcal P}_{1,0}$ when predicting Call option prices using the ANN with soft constraints.}
\label{arbTab}
\end{center}
\end{table}
It can be seen that training with no arbitrage constraints produces a high arbitrage penalty both in-sample and out-of sample\footnote{The in-sample arbitrage is bigger because the number of in-sample inputs is 182016, while the number of out-of sample inputs is 4 times less.}. This justifies our hypothesis that by default the ANN doesn't support no-arbitrage of the option prices.  Increasing the penalty barrier by increasing all $\lambda$ coefficients decreases the arbitrage and almost eliminates it. This is, however, obtained by the cost of the minor increase of the MSE, which is the usual case for the constraint optimization. The results for the last line in Table~\ref{arbTab} are also presented in Fig.~\ref{BSpriceInNA},\ref{BSpriceOutNA} for in-sample and out-of sample inputs. And Fig.~\ref{BSdenInNA}, \ref{BSdenOutNA} represent the density of the difference $\delta = y - y_{\ann}$. The plots demonstrate that the no-arbitrage conditions deviate deep ITM prices predicted by the ANN from the actual prices, while they practically don't affect the prices for the other moneynesses. As we are using soft constraints, even at $\lambda = 100$ the arbitrage is not fully eliminated (in contrast to hard constraints), but is reduced to almost a negligible level.

Certainly, the elapsed time necessary to train this network increases, e.g., in our example for the last line in Table~\ref{arbTab} it is 40-60 secs per epoch. Also, we need more epochs to achieve better convergence. Therefore, in this test the number of epochs was increased to 30.

As the result, thus trained ANN guarantees no-arbitrage prices in-sample. However, it doesn't guarantee arbitrage-free prices out-of sample. We leave a general solution of this problem as an open question for a moment. However, in some particular cases it is possible to provide a recipe. For instance, suppose we are using a ANN approach to build an implied volatility surface given market quotes and constant input parameters $\bm{\theta} =[S,r,q, \sigma]$. Then, given some market quote $C(T, K)$ for the {\it out-of sample} pair $(T, K) $, we can find the closest to $K$ {\it in-sample} pairs $(T, K_1, K_2, K_3), \ K_1 < K < K_3, \ K_1 < K_2 < K_3$, and get the corresponding prices $C_i = C_\ann(\bm{\theta}, T, K_i), \ i= 1,2,3$. Next, some version of no-arbitrage interpolation could be used that given the values of $K_i, C_i, \ i=1,2,3$ provides an arbitrage-free value $C_{\af}(T, K)$, see, e.g., \cite{ItkinLipton2017,ELVG}. And then this value should be used instead of $C_\ann(\bm{\theta}, T, K)$. In case when such in-sample pairs are not available, or belong to different maturities, construction of no-arbitrage interpolation becomes more cumbersome.

\section{Calibration of option pricing models using ANN}

According to \cite{Cont2010}, a derivative pricing model is said to be calibrated to a set of benchmark instruments if the values of these instruments, computed in the model, correspond to their market prices. Model calibration is the procedure of selecting model parameters in order to verify the calibration condition. Model calibration can be viewed as the inverse problem associated with the pricing of derivatives. In the theoretical situation where prices of Call options are available for all strikes and maturities, the calibration problem can be explicitly solved using an inversion formula. In real situations, given a finite (and often sparse) set of derivative prices, model calibration is an ill‐posed problem whose solution often requires a regularization method.

Mathematically this is equivalent to solving an optimization problem. Suppose we are given a set of Call options market prices $C_{M}$ for a set of maturities $T_i, \ i=1,..,n$ and strikes $K_{i,j}, \ j = 1,...,m_i$ corresponding to the known market input data such as $S,r,q$, etc. This notation implies that the number of traded strikes $K_{i,j}$ could differ for each maturity $T_i$. Suppose we calibrate model $\calM$ described in Section~\ref{model} which has the model parameters $\bm{p}$ and the input data parameters $\bm{\theta}$. We differentiate $\bm{p}$ and $\bm{\theta}$ by their notion: parameters $\bm{p}$ have to be found by calibration while the input data $\bm{\theta}$ represent some observable (or known) market values, like $S,r,q,T,K$\footnote{Here as an example we discuss just the simplest case. In more complex models $r,q$ could be stochastic variables with their own models which are characterized by additional set of parameters $\bm{p}$.}.

Given the values of $\bm{\theta}$ and $\bm{p}$, the model $\calM$ can generate the theoretical (model) Call option prices  $C$. Then the following optimization problem needs to be solved to obtain the calibrated parameters of the model:
\begin{equation}  \label{opt}
\argmin_{\bm{p}} \sum_{i=1}^n \sum_{j=1}^{m_i} \omega_{i,j} \| C_M(\bm{\theta}_{i,j}) -
 C(\bm{\theta}_{i,j}, \bm{p})\|
 \end{equation}
\noindent where $\omega_{i,j}$ is a set of weights, and a typical norm $\|\cdot\|$ used in \eqref{opt} is $L^2$. In general, solving this problem for the non-linear function  $C(\bm{\theta}_{i,j}, \bm{p})$ requires using algorithms of global optimization, as the multi-dimensional objective (loss) function might have a lot of local minima. Also, this problem could be constrained, which, e.g., is a typical case when the implied volatility surface is constructed based on the market quotes, see again \cite{ItkinSigmoid2015} and references therein.

Since the pricer computing $C(\bm{\theta}_{i,j}, \bm{p})$ could be slow, e.g., when it uses Monte Carlo simulation, solving this optimization problem could be extremely time-consuming, while frequent re-calibration would be desirable for front offices and trading desks. Therefore, having in mind a significant recent progress of DL and ANN, it was proposed to replace the prices  $C(\bm{\theta}_{i,j}, \bm{p})$ with their ANN approximation as this was discussed in above. A nice survey of this approach could be found in \cite{Horvath2019, Kees2019}, therefore, we refer the reader to those papers. Since the ANN can be trained offline, and getting the prices from the trained ANN is very fast, the first bottleneck in solving the optimization problem disappears. However, in both \cite{Horvath2019, Kees2019} the global optimization problem is still solved by using traditional numerical methods.

In our opinion this second step (global optimization) could be fully eliminated by using the inverse map  approach. This approach was, in particular, considered in \cite{Hernandez2017} where the network is trained to directly return calibrated parameters of the stochastic model. It was, however, reported that such an approach is unstable or doesn't converge. A similar behavior was confirmed by \cite{Tomas2019}. Therefore, e.g., in \cite{Horvath2019} the choice is made in favor of a two-step process. The authors claim that separation of building the approximation ANN and the calibration brings the following improvements: i) the training becomes more robust (with respect to generalisation errors on unseen data, and ii) the speed-up of the online calibration part is even more pronounced than in direct calibration to data. The latter argument, however, is not favorable as compared with the inverse map approach, as for the online (out-of sample) calibration step the latter approach also uses a previously trained ANN. This is exactly same what the cited authors do. But the former argument requires a special discussion which is provided in Section~\ref{robust}.

Below we propose another DL approach to calibrate the model with no second step (global optimization). This approach is constructed based on the inverse map, but differs in the last steps.

\subsection{The inverse map approach}

As the first example consider again a simple Black-Scholes framework where the only parameter of the model is the implied volatility $\sigma$. In Section~\ref{ANNGreeks} we generated random vectors $\bm{\xi} = [S,K,T,r,q,\sigma]$, and then computed the Call option price $C$ using the Black-Scholes model. In other words, by doing so we constructed a map $\calM_d\from [\bm{p},\bm{\theta}] \mapsto C$.  However, once the input  $\bm{\xi}  = [S,K,T,r,q,\sigma]$ and the output $C(S,K,T,r,q,\sigma)$ are generated, these variables could be re-arrange to produce the inverse map $\calM_i\from [S,K,T,r,q,C] \mapsto \sigma$, or, in general, $\calM_i\from [\bm{\theta},C] \mapsto \bm{p}$. Then, by using the same procedure as described in Section~\ref{ANNGreeks} we build a feedforward ANN. Given a vector  $\bm{\xi} = [S,K,T,r,q,C]$ this ANN returns the implied volatility $\sigma$.

A similar (despite, perhaps, more general, approach) is discussed in \cite{INN2018}, where  a particular class of neural networks is considered to be well suited for this task – so-called Invertible Neural Networks (INNs). The authors verify experimentally, on artificial data and real-world problems from astrophysics and medicine, that INNs are a powerful analysis tool to find multi-modalities in parameter space, to uncover parameter correlations, and to identify unrecoverable parameters. Therefore, application of INNs to the calibration problem in finance could be an alternative approach.

The inverse map $\calM_i\from [\bm{\theta},C] \mapsto \bm{p}$ can be used for calibration of a given model. However, it could not be used directly. Indeed, suppose again that we are given a set of Call options market prices $C_{M}$ for a set of maturities $T_i, \ i=1,..,n$ and strikes $K_{i,j}, \ j = 1,...,m_i$ corresponding to the known market input data $\bm{\theta} = [S,r,q]$. Then, for each particular set $\bm{\xi}_{i,j} = [\bm{\theta}, T_i, K_{i,j},  C(\bm{\theta},T_i, K_{i,j}, \sigma_{i,j})]$ being used as an input to the ANN, the latter will return the unique value $\sigma_{\ann}(\bm{\xi}_{i,j})$. Thus, the trained ANN doesn't solve the calibration problem which requires a single value of the parameter $\sigma$ to solve \eqref{opt}.

Therefore, it is useful to reformulate \eqref{opt} in the following form
\begin{equation}  \label{opt1}
\argmin_{\sigma_\calM} \sum_{i=1}^n \sum_{j=1}^{m_i} \bar{\omega}_{i,j} \| \sigma_{\ann} \left(\bm{\xi}_{i,j}\right) -  \sigma_\calM \|.
 \end{equation}
\noindent where $\sigma_\calM$ denotes the calibrated value of the implied volatility, and $\bar{w}$ are some weights. In a general case of the model with $N$ parameters $\bm{p} = p_1,...,p_N$, this could be re-written as
\begin{equation}  \label{optN}
\argmin_{\bm{p} } \sum_{i=1}^n \sum_{j=1}^{m_i} \sum_{k=1}^{N}
\bar{\omega}_{i,j,k} \| p_{k,\ann} \left(\bm{\xi}_{i,j}\right) -  p_k\|.
 \end{equation}
This is also an optimization problem, but under $L^2$ norm it can be solved analytically. The obvious solution read
\begin{equation} \label{sol}
p_k = \frac{\sum_{i=1}^n \sum_{j=1}^{m_i} \bar{\omega}_{i,j,k} p_{k,\ann} \left(\bm{\xi}_{i,j}\right)}{\sum_{i=1}^n \sum_{j=1}^{m_i} \bar{\omega}_{i,j,k} }.
\end{equation}
Thus, the algorithm of finding the calibrated values of all the parameters $\bm{p}$ also consists of two steps. At the first we use the  inverse map $\calM_i$ and create a trained feedforward ANN. At the second step we substitute the market data into the ANN to obtain $\bm{p}_{\ann}(\bm{\xi}_{i,j})$, and then use \eqref{optN} to find the calibrated values of $\bm{p}$. However, it doesn't require solving any optimization problem as this problem is already solved analytically.

A valid question to ask would be: how the objective function in \eqref{optN} differs from that in \eqref{opt}? Do we solve the same problem, or this statement of the calibration problem is quite different? To answer this, first observe that the approach in \cite{Horvath2019, Kees2019} replaces \eqref{opt} with another problem
\begin{equation}  \label{optANN}
\argmin_{\bm{p}} \sum_{i=1}^n \sum_{j=1}^{m_i} \omega_{i,j} \| C_M(\bm{\theta}_{i,j}) -
 C_{\ann}(\bm{\theta}_{i,j}, \bm{p})\|
 \end{equation}
 Based on the direct map $\calM_d\from [\bm{p},\bm{\theta}] \mapsto C$ generated by the trained ANN we can re-write this in the form
\begin{align}  \label{optANN2}
&\ \  \argmin_{\bm{p}} \sum_{i=1}^n \sum_{j=1}^{m_i} \omega_{i,j} \| C_{\ann}(\bm{\theta}_{i,j}, \bm{p}_{i,j}) -  C_{\ann}(\bm{\theta}_{i,j}, \bm{p})\| \\
 &= \argmin_{\bm{p}} \sum_{i=1}^n \sum_{j=1}^{m_i} \omega_{i,j} \| C_{\ann}\left(
 \bm{\theta}_{i,j}, \bm{p}_{\ann}(\bm{\xi}_{i,j}) \right) -  C_{\ann}(\bm{\theta}_{i,j}, \bm{p})\|,  \nonumber
 \end{align}
 \noindent where $[\bm{\theta}_{i,j}, \bm{p}_{i,j}]$ is the vector of input data and parameters of the model which  being given as the input to the ANN returns the market price  $C_M(\bm{\theta}_{i,j})$. Now expanding \eqref{optANN} into Taylor series on all parameters $\bm{p}$ we obtain
\begin{align}  \label{optANN3}
C_{\ann}(\bm{\theta}_{i,j}, \bm{p}_{i,j}) &-  C_{\ann}(\bm{\theta}_{i,j}, \bm{p}) \\
&= \sum_{k=1}^N \fp{ C_{\ann}(\bm{\theta}_{i,j}, \bm{p}_{i,j}) }{p_{k,\ann}(\bm{\xi}_{i,j})} \left[p_{k,\ann}(\bm{\xi}_{i,j}) - p_k\right] + O\left( (p_{k,\ann}(\bm{\xi}_{i,j}) - p_k)^2 \right). \nonumber
\end{align}
And then in the first order of approximation on each $\Delta p_{i,j,k} = p_{k,\ann}(\bm{\xi}_{i,j}) - p_k$ under the norm $|\cdot|$, we obtain
\begin{align} \label{major}
\sum_{i=1}^n \sum_{j=1}^{m_i} & \omega_{i,j} | C_M(\bm{\theta}_{i,j}) -
 C_{\ann}(\bm{\theta}_{i,j}, \bm{p})|  = \sum_{i=1}^n \sum_{j=1}^{m_i} \omega_{i,j}
\left|\sum_{k=1}^N \fp{ C_{\ann}(\bm{\theta}_{i,j}, \bm{p}_{i,j}) }{p_{k,\ann}(\bm{\xi}_{i,j})} \left(p_{k,\ann}(\bm{\xi}_{i,j}) - p_k\right) \right| \\
 & \le \sum_{i=1}^n \sum_{j=1}^{m_i} \omega_{i,j} \sum_{k=1}^N
\left|\fp{ C_{\ann}(\bm{\theta}_{i,j}, \bm{p}_{i,j}) }{p_{k,\ann}(\bm{\xi}_{i,j})} \right| \left|p_{k,\ann}(\bm{\xi}_{i,j}) - p_k \right|
= \sum_{i=1}^n \sum_{j=1}^{m_i} \sum_{k=1}^N \bar{\omega}_{i,j,k} \left|p_{k,\ann}(\bm{\xi}_{i,j}) - p_k \right|, \nonumber \\
\bar{\omega}_{i,j,k} &= \omega_{i,j} \left|\fp{ C_{\ann}(\bm{\theta}_{i,j}, \bm{p}_{i,j}) }{p_{k,\ann}(\bm{\xi}_{i,j})} \right|. \nonumber
\end{align}
All sums in the first two lines of \eqref{major} are non-negative and depend on the same set of parameters $\bm{p}$. Therefore, the solution of the problem
\begin{equation} \label{probP}
\argmin_{\bm{p}} \sum_{i=1}^n \sum_{j=1}^{m_i} \sum_{k=1}^N \bar{\omega}_{i,j,k} \left|p_{k,\ann}(\bm{\xi}_{i,j}) - p_k \right|
\end{equation}
\noindent also solves the problem
\begin{equation} \label{probC}
\argmin_{\bm{p}} \sum_{i=1}^n \sum_{j=1}^{m_i} \omega_{i,j} | C_M(\bm{\theta}_{i,j}) -
 C_{\ann}(\bm{\theta}_{i,j}, \bm{p})|,
\end{equation}
\noindent assuming that all $\bar{\omega}_{i,j,k}$ exist and are finite. Since $|x| = \sqrt{x^2}$, the solution $\bm{p}$ of \eqref{probP} given by \eqref{sol} also solves \eqref{probC} up to $O(\sum_{k,i,j} (\Delta p_{i,j,k})^2)$.

The derivatives of the Call option price on each parameter, as defined in the last line of \eqref{major}, are computed by the ANN automatically (by using AAD), and, thus, are known. Therefore, given weights $w_{i,j}$ all weights $\bar{w}_{i,j,k}$ can be obtained at each node of the ANN for free.

It is important to underline that, in order to be consistent when switching from \eqref{optANN} to \eqref{optANN2}, we use the following sequence of steps:
\begin{enumerate}
\item Given a set of inputs $\bm{\xi}$ we use the model pricer to generate a set of corresponding model outputs $C(\bm{\xi})$.
\item We use all pairs $\{\bm{\xi},  C(\bm{\xi})\}$ to generate a map $\calM_d\from \bm{\xi} \mapsto C_{\ann}(\bm{\xi})$. This map is produced by training the ANN, i.e. by running a constraint optimization, and so taking into account no-arbitrage conditions in-sample.
\item We now re-arrange variables to produce the inverse map $\calM_{i, \ann}\from \{\bm{\theta},  C_{\ann}(\bm{\xi})\} \mapsto \bm{p}$. It is worth mentioning that, in general, this map differs from the map $\calM_i\from \{\bm{\theta},  C(\bm{\xi})\} \mapsto \bm{p}$.
\item Having the inverse map we use it together with the market data to generate outputs $p_{k,\ann}((\bm{\xi}_{i,j})$, and then \eqref{sol} together with the last line of \eqref{major}.

\end{enumerate}

In more sophisticated calibrations a part of the input data parameters could also change. For instance, the option market data could reflect not just a current snapshot at time $t$, but also some historical data at $\tau < t$, where the stock price $S_\tau \ne S_t$, etc. Then, this extends a set of indexes $i,j$ to reflect the dependence of the Call option price on $S$, but this extension is straightforward.

\subsection{Stability and convergence} \label{robust}

Here again we consider the simple example described in Section~\ref{ANNGreeks}, which uses the Black-Scholes model. Now we want to build an inverse map and solve the problem of calibration of the implied volatility $\sigma$ given the market option prices. The only difference as compared with the direct approach is that after all training sets are generated, we switch the Call price $C$ and $\sigma$, thus producing a map $\calM_i\from [S,K,T,r,q,C] \mapsto \sigma$. Next, if then we train the ANN using this data as inputs and outputs in the same way how this is done in Section~\ref{ANNGreeks} (by using the same architecture of the ANN and even the same code),  the obtained results look discouraged. Training is definitely unstable and of poor quality even for in-sample data, thus, in agreement with what was found in \cite{Hernandez2017,Tomas2019}.

\begin{figure}[!htb]
\captionsetup{format=plain}
\begin{center}
\includegraphics[totalheight=0.4\linewidth]{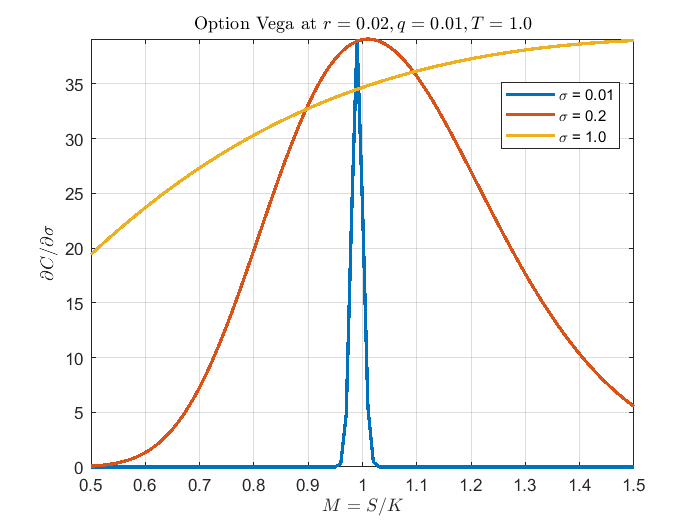}
\caption{Option Vega $\fp{C}{\sigma}$ in a test example for the Black-Scholes model.}
\label{gradSigC}
\end{center}
\end{figure}

This, however, could be naturally explained. As the option price $C$ is now an input parameter, and $\sigma$ is an output, consider the gradient $\fp{\sigma}{C} = 1/\mathrm{Vega}$, where $\mathrm{Vega}$ is the option Vega. The latter can be easily computed using the Black-Scholes formula, and the well-known behavior of Vega is presented in Fig.~\ref{gradSigC} for $S \in [50,150], K=100, r = 0.02, q = 0.01, T = 1.0$ and three volatilities $T=0.01,0.2,1$.
Obviously, Vega is a bell-shaped function. When the product $\kappa = \sigma \sqrt{T}$ is small, the inverse gradient is very high at small and high moneyness $M$, and  small At-The-Money (ATM). This creates a problem for training the ANN which is known in the literature as the Exploding Gradients problem, This problem already got attention of the society, see, e.g., \cite{NNMethods2017}. Various methods to deal with it have been proposed including: reducing the batch size, changing the architecture of the ANN by switching to Long Short-Term Memory Network, having fewer hidden layers of the network, etc. Also, if exploding gradients occur, one can check for and limit the size of gradients during the training of the ANN. This is called gradient clipping. As per \cite{NNMethods2017}, dealing with the exploding gradients has a simple but very effective solution: clipping gradients if their norm exceeds a given threshold. In particular, Keras library provides an interface for setting the clipnorm or clipvalue arguments on your favorite optimizer before training.

To choose the value of the clipnorm we utilized the following approach. First we define a callback function by using LearningRateScheduler class of Keras
\begin{lstlisting}[basicstyle=\scriptsize]
lrs = LearningRateScheduler(my_callback)
\end{lstlisting}
Then inside this callback after the first epoch we compute the norm by using
\begin{lstlisting}[basicstyle=\scriptsize]
norm = np.sqrt(sum(np.sum(K.get_value(w)) for w in model.optimizer.weights))
\end{lstlisting}
Based on that we set the initial value of clipnorm. This value is dynamically updated, if necessary, by another custom callback function.

Another important thing is scaling. Scaling the inputs first, makes sense if one reveals some symmetries (like in the Black-Scholes model for the European vanilla options). In this case scaling allows reduction of the number of the input variables. Second, good scaling of both inputs and outputs helps with changing the norm of the problem gradients. Therefore, it also helps in solving the Exploding Gradients problem. In particular, as applied to our test, we replace the output variable $\sigma$ with
\[ \sigma \mapsto N\left(\frac{\log(S/K)}{\sigma \sqrt{T}}\right) \equiv \calN(\sigma), \]
\noindent where $N(x)$ is the normal CDF. Thus, all outputs are in $[0,1]$. This scaling is chosen for reasons mentioned in above, and also because the inverse map $\calN(\sigma) \mapsto \sigma$ could be done analytically. Other functions could also be proposed and work well, e.g.,
\[ \sigma \mapsto \frac{1}{2}[1 + \tanh\left(\frac{\log(S/K)}{\sigma \sqrt{T}}\right)], \]
\noindent   etc.

We used the same ANN design as in Section~\ref{ANNGreeks} with the optimizer set to Adam.
We also use another Keras callback ReduceLROnPlateau. Models often benefit from reducing the learning rate by a factor of 2-10 once learning stagnates. This callback monitors a quantity, and if no improvement is seen for a "patience" number of epochs, the learning rate is reduced.

\begin{figure}[H]
\captionsetup{format=plain}
\begin{minipage}{0.4\textwidth}
\centering
\hspace*{-0.2in}
\includegraphics[totalheight=2.2in]{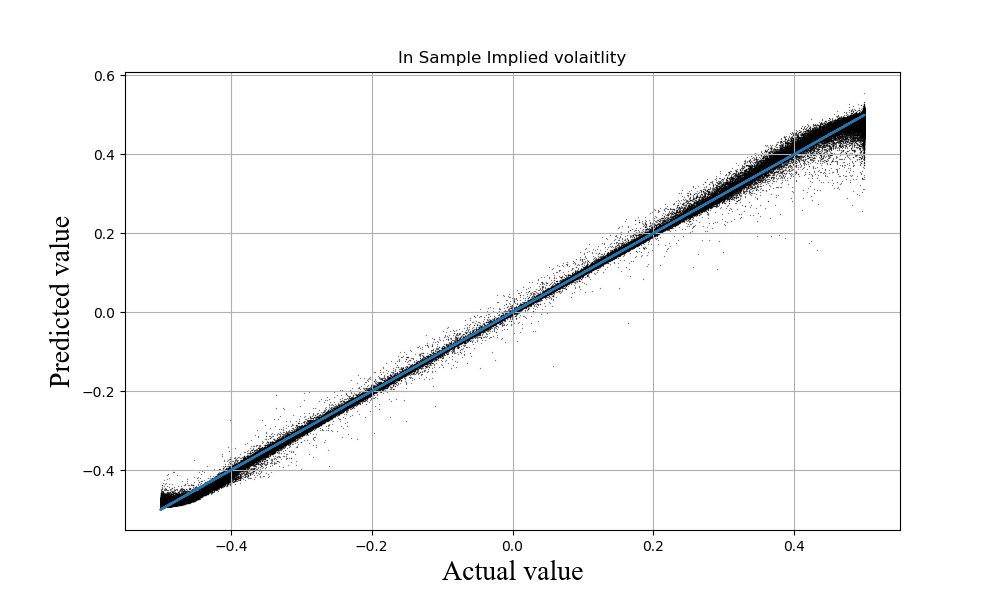}
\caption{In-sample ANN Black-Scholes scaled implied volatility $\calN(\sigma)$ vs the Black-Scholes $\calN(\sigma)$.}
\label{BSivIn}
\end{minipage}
\hspace{0.1\textwidth}
\begin{minipage}{0.4\textwidth}
\centering
\hspace*{-0.2in}
\includegraphics[totalheight=2.2in]{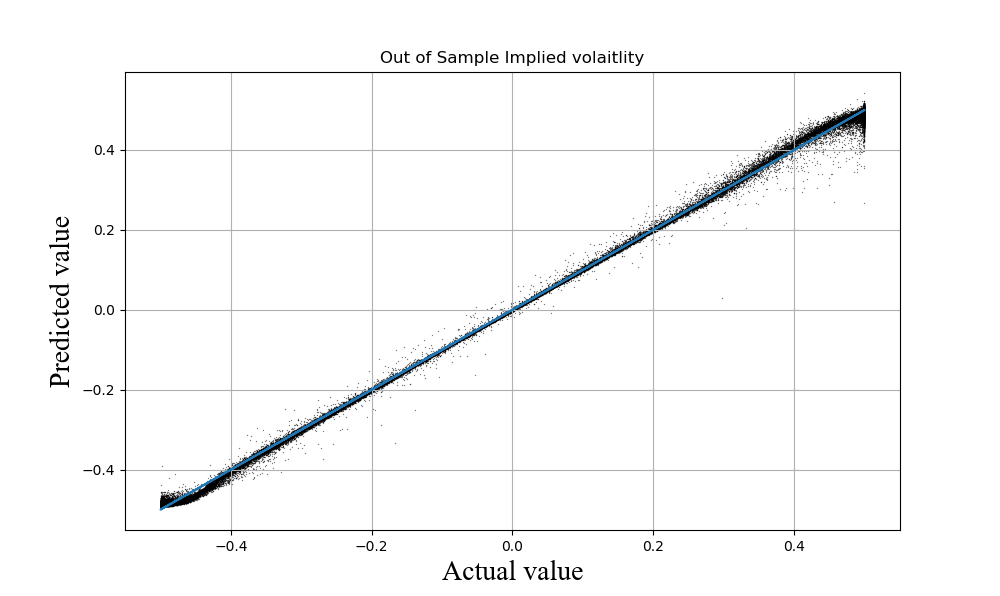}
\caption{Out-of sample ANN Black-Scholes scaled implied volatility $\calN(\sigma)$ vs the Black-Scholes $\calN(\sigma)$.}
\label{BSivOut}
\end{minipage}
\end{figure}

\begin{figure}[H]
\captionsetup{format=plain}
\begin{minipage}{0.4\textwidth}
\centering
\hspace*{-0.2in}
\includegraphics[totalheight=2.2in]{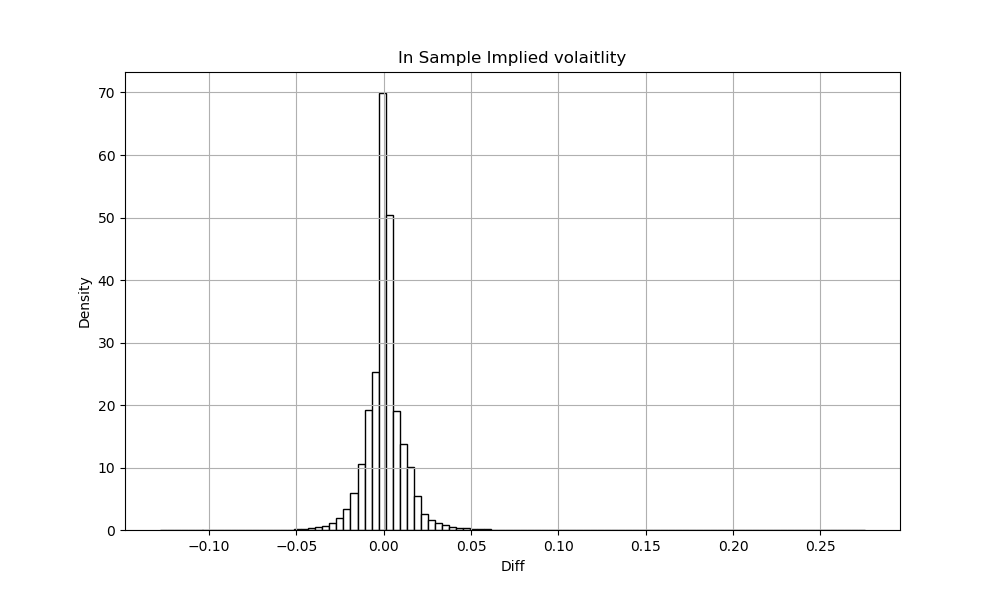}
\caption{In-sample $\delta$ error density of the Black-Scholes $\calN(\sigma)$.}
\label{BSdenIVIn}
\end{minipage}
\hspace{0.1\textwidth}
\begin{minipage}{0.4\textwidth}
\centering
\hspace*{-0.2in}
\includegraphics[totalheight=2.2in]{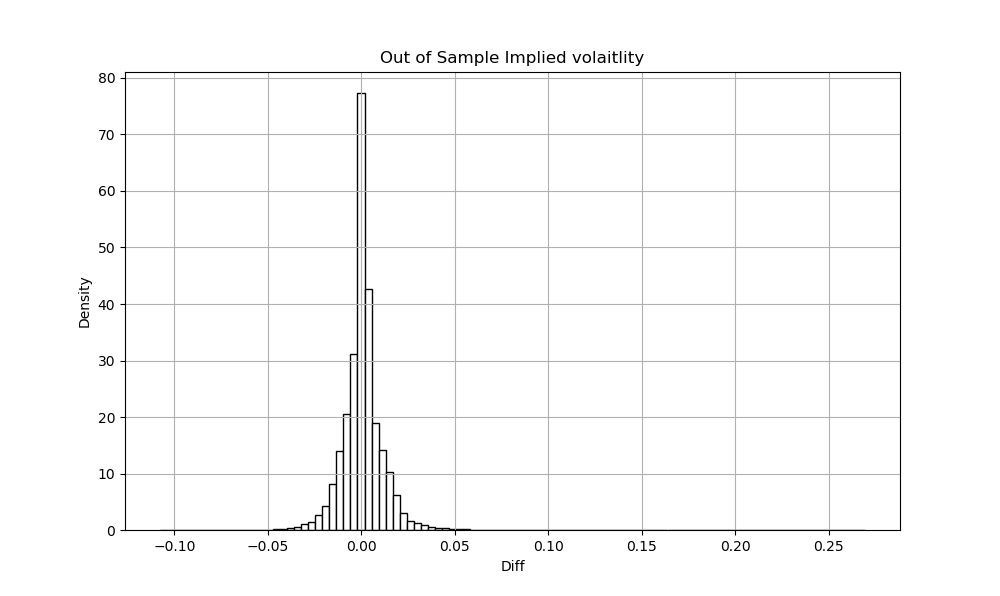}
\caption{Out-of sample $\delta$ error density of the Black-Scholes $\calN(\sigma)$.}
\label{BSdenIVOut}
\end{minipage}
\end{figure}

 The results of this test are presented in Fig.~\ref{BSivIn},\ref{BSivOut} for in-sample and out-of sample inputs. And Fig.~\ref{BSdenIVIn},\ref{BSdenIVOut} represent the density of the difference $\delta = y - y_{\ann}$. The MSE both in-sample and out-of sample is 1.5 bps, and Mean percent error is -0.3 after 30 epochs.

Our experiments show that the training of the ANN is stable and converges. Still, some improvements could be desirable at very small and high implied volatilities, where, as known, the Call option price reaches its intrinsic value, and, thus, the inverse map $\calM_i\from C \mapsto \sigma$ is not well-defined at the fixed machine arithmetic.

\section*{Conclusions}

In this paper we discuss calibration of option pricing models to market data by using the DL approach, as it was recently drawn attention of the financial society. We highlight some pitfalls in the existing approaches and propose resolutions that improve both performance and accuracy of calibration. In particular, we advocate the following improvements: i) a choice of the activation functions to be $C^2$; ii) no-arbitrage conditions to be taken into account as penalties when training the pricing ANN; direct calibration of the model parameters by using the inverse map approach. The latter is further extended by taking into account no-arbitrage. We show how the classical calibration problem can be re-formulated by using the inverse map. We also discuss some technical issues inherent to this approach and reported in the literature, and show several ways to overcome them. Simple examples provided for the Black-Scholes model as applied to European Call options pricing and calibration problems demonstrate the robustness of the proposed method.

Further applications of the method include numerical experiments with more sophisticated models with a number of model parameters (outputs). These results will be reported elsewhere.


\section*{Acknowledgments}
I thank Peter Carr and Igor Halperin for thoughtful comments, and participants of QuantMinds 2019 International Conference for useful discussions, although any errors are my own.

\vspace{0.3in}
\bibliographystyle{elsarticle-harv}
\newcommand{\noopsort}[1]{} \newcommand{\printfirst}[2]{#1}
  \newcommand{\singleletter}[1]{#1} \newcommand{\switchargs}[2]{#2#1}

\end{document}